\documentclass[useAMS]{mn2e}
\usepackage{epsfig}

\title[The formation of the extremely primitive star 
 relies on dust]{ The formation of the extremely primitive star 
SDSS J102915+172927 relies on dust}
\author[R. Schneider et al.]{Raffaella Schneider$^{1}$\thanks{E-mail:
raffaella.schneider@oa-roma.inaf.it}, Kazuyuki Omukai$^{2}$, Marco Limongi$^{1}$, Andrea Ferrara$^3$, 
\newauthor
Ruben Salvaterra$^4$, Alessandro Chieffi$^5$, Simone Bianchi$^6$\\
$^{1}$INAF/Osservatorio Astronomico di Roma, Via di Frascati 33, 00040 Monteporzio, Italy \\
$^{2}$Department of Physics, Kyoto University, Kyoto 606-8502, Japan\\
$^{3}$Scuola Normale Superiore, Piazza dei Cavalieri 7, 56126 Pisa, Italy\\
$^{4}$INAF/IASF, Via Bassini 15, 20133 Milano, Italy\\
$^{5}$INAF/IASF, Via Fosso del Cavaliere 100, 00133 Roma, Italy\\
$^{6}$INAF/Osservatorio Astrofisico di Arcetri, Largo Enrico Fermi 5, 50125 Firenze, Italy}

\begin{document}

\date{}

\pagerange{\pageref{firstpage}--\pageref{lastpage}} \pubyear{2012}

\maketitle

\label{firstpage}

\begin{abstract}
The relative importance of metals and dust grains in the formation of the first low-mass
stars has been a subject of debate. 
The recently discovered Galactic halo star SDSS J102915+172927 
(Caffau et al. 2011) has a mass less than $\rm 0.8~M_{\odot}$ and a metallicity of $Z = 4.5 \times 10^{-5} Z_{\odot}$.
We investigate the origin and properties of this star by reconstructing the physical conditions in its
birth cloud. We show that the observed elemental abundance trend of SDSS J102915+172927 can be well fitted by
the yields of core-collapse supernovae with metal-free progenitors of $\rm 20~M_{\odot}$ and $\rm 35~M_{\odot}$.
Using these selected supernova explosion models, we compute the corresponding dust yields and the resulting dust
depletion factor taking into account the partial destruction by the supernova reverse shock. We then follow the
collapse and fragmentation of a star forming cloud enriched by the products of these SN explosions at the observed
metallicity of SDSS J102915+172927. 
We find that $\rm [0.05 - 0.1]~M_{\odot}$ mass fragments, which then lead to the formation of low-mass stars, 
can occur provided that the mass fraction of dust grains in the birth cloud exceeds 0.01 of the total mass of 
metals and dust. This, in turn, requires that at least $\rm 0.4~M_{\odot}$ of dust condense in the first supernovae, 
allowing for moderate destruction by the reverse shock. 
If dust formation in the first supernovae is less efficient or strong dust destruction does occur, 
the thermal evolution of the SDSS J102915+172927 birth cloud is dominated by molecular cooling, and only
$\rm \ge 8~M_{\odot}$ fragments can form. 
We conclude that the observed properties of SDSS J102915+172927 support the suggestion that dust must have
condensed in the ejecta of the first supernovae and played a fundamental role in the formation of the first low-mass stars.
\end{abstract}


\section{Introduction}
The observed elemental abundances of metal-poor stars in the halo of the Milky Way
or in its dwarf satellites record the physical conditions prevailing in star forming regions
at high redshift. The mass fraction of Fe relative to H, [Fe/H]$\rm = Log(M_{Fe}/M_{H}) - 
Log(M_{Fe}/M_{H})_{\odot}$, normalized to the abundance in the Sun, is generally used
as a metallicity tracer. Galactic halo stars show a remarkably uniform trend of
elemental abundance ratios (Cayrel et al. 2004) at [Fe/H]$ < -2.5$. An exception are the so-called
carbon-enhanced extremely metal-poor stars (Beers \& Christlieb 2005), 
CEMP, which show a large C over-abundance, [C/Fe]$ > 1$. 
Interestingly, the fraction of carbon-enhanced stars
increases with decreasing metallicity and 3 out of the 4 known halo stars with [Fe/H]$ < -4$ are
CEMPs (Christlieb et al. 2002; Frebel et al. 2005; Norris et al. 2007).
Since CII and OI fine structure lines are amongst the most efficient gas coolants (Bromm \& Loeb 2003), 
the observed CNO enhancement has been interpreted as the key environmental condition for the formation of
low-mass stars (Frebel, Johnson \& Bromm 2007). This hypothesis has been recently challenged by the discovery of
SDSS J102915+172927 (Caffau et al. 2011), which has [Fe/H]$ = -4.99$ and a chemical pattern typical of classical
extremely metal-poor stars, i.e. no CNO enhancement; its inferred metallicity falls in the critical range,
$Z_{\rm cr} = 10^{-5 \pm 1} Z_{\odot}$, in which low-mass star formation is possible provided that a
fraction of the metals is in the form of dust grains (Schneider et al. 2002, 2003; Omukai et al. 2005; 
Schneider et al. 2006; Schneider et al. 2011).

In this work, we investigate the origin of SDSS J102915+172927 by reconstructing the physical conditions in its birth
environment. We use detailed pre-supernova and supernova (SN) explosion models to fit the observed elemental abundances
of SDSS J102915+172927 and identify plausible supernova progenitors responsible for the enrichment of its birth cloud 
(Section 2). 
We use these supernova explosion models to calculate the amount of dust produced in their ejecta and its partial destruction
by the associated reverse shock (Section 3). We then study the collapse and fragmentation properties of 
a star forming cloud enriched by the products of these explosions and identify the conditions that enable the formation
of sub-solar mass stars (Section 4). Finally, in Section 5 we draw our conclusions.

\section{SDSS J102915+172927 elemental abundances fit}
The observed elemental abundances of extremely metal-poor stars (EMPs, with [Fe/H] $\le -3$) 
and ultra metal-poor stars (UMPs, with [Fe/H] $\le -4$) have been interpreted in terms of 
individual metal-free SN explosion models (Umeda \& Nomoto 2003; Iwamoto et al. 2005;
Tominaga, Umeda \& Nomoto 2007) or involving a combination of metal-free SN models with different progenitor masses, 
explosion energies, and mixing (Limongi et al. 2003; Heger \& Woosley 2010). 
Samples of EMPs with typical halo signatures (Cayrel et al. 2004) can be well reproduced by the 
yields of Population III (Pop III) massive ($\rm 30 - 50~M_{\odot}$) energetic hypernovae (Tominaga et al. 2007) or by
the integrated yields of ordinary Pop III SN with reduced mixing and explosion energy (Heger \& Woosley 2010). The
peculiar abundance pattern of extreme CEMPs such as HE0107-5240 with [Fe/H] = -5.2 (Christlieb et al. 2002) and 
HE1327-2326 with [Fe/H] = -5.4 (Frebel et al. 2005) has been interpreted with the yields of a Pop III 25~M$_\odot$
faint SN that underwent strong mixing and fallback (Umeda \& Nomoto 2003; Iwamoto et al. 2005), with the pollution
of the birth cloud by two core-collapse Pop III SN (Limongi, Chieffi \& Bonifacio 2003), or with the yields of Pop III SN with 
progenitor mass $\rm 12 - 30~M_{\odot}$ with reduced mixing and explosion energy (Heger \& Woosley 2010).
Additional hypothesis involve mass transfer of CNO elements from a postulated binary companion (Suda et al. 2004)
or mass loss by massive ($\rm 60~M_{\odot}$) near metal-free stars (Meynet, Ekstr\"{o}m \& Maeder 2006). 
It is clear from the above, that a solid picture for the origin of EMPs and, in particular for CEMPs, is still lacking.

Despite its very low iron content, the recently discovered halo star SDSS J102915+172927 shows an abundance 
pattern consistent with typical Galactic halo signatures. Our aim here is to identify a set of plausible Pop III
SN models which may have polluted its birth cloud. The procedure that we follow is to fit the 
elemental abundance patter of SDSS 102915+172927 with the chemical yields produced by a new set of Pop III 
core-collapse SN models. We refer the interested reader to Limongi \& Chieffi (2012) for a thorough presentation
of the models. Here we simply recall the basic properties that are relevant for the present investigation.

The set of models extends in mass between 13 and 80 $\rm M_\odot$ and has been followed from the pre-main sequence phase up 
to the onset of the iron core collapse by means of the FRANEC stellar evolutionary code. The initial composition is set to the 
pristine Big Bang nucleosynthesis one (Y=0.23) and mass loss is switched off during 
all the evolutionary stages. Calculations based on the radiatively driven wind theory show that the mass loss scales 
with the metallicity of the surface of the star as $(Z/Z_\odot)^{\alpha}$ with $\alpha$ between 0.5 and 0.8 
(Kudritzki et al. 1987; Vink et al. 2001). Such a dependence would become much stronger at lower $Z$ (Kudritzki 2002).
Hence we expect mass loss to be strongly inhibited in zero metallicity stars (but see also Meynet et al. 2006). 
Internal mixing induced by rotation or mass transfer from a companion star could modify the surface chemical composition, 
enhancing the fraction of metals in the atmosphere. Efficient mass loss would affect the yields of the light elements 
(up to Al at most) but would not change the yields of the iron group elements, because these mostly depend on the 
dynamics of the explosion.


The explosive nucleosynthesis has been computed in the framework of the induced explosion 
(Limongi \& Chieffi 2006) approximation. 
More specifically, the explosion of the mantle of the star is started by imparting instantaneously 
an initial velocity $v_0$ to a mass coordinate of $\rm \sim 1~M_\odot$ of the pre-supernova model, 
i.e., well within the iron core. The propagation of the shock wave that forms consequently is 
followed by means of an hydro code that solves the fully compressible reactive hydrodynamic 
equations using the piecewise parabolic method (PPM) in the Lagrangean form 
(see Chieffi \& Limongi 2002 and Limongi \& Chieffi 2003). 

Since the energy of the shock wave that drives the ejection of part of the collapsing star can not 
yet be determined on the basis of first principles, some kind of $\it arbitrary$ choice is necessary. 
The standard procedure followed by most of the leading groups working in this field
(see, e.g., Umeda \& Nomoto 2003, 2005; Tominaga et al. 2007; Heger \& Woosley 2010) is
to calibrate the explosion by requiring the final kinetic energy of the ejecta to have a 
specific value, the ejection of a given amount of $\rm ^{56}Ni$, 
or by choosing the mass of the remnant in order to fit a specific elemental ratio.
Here we follow a slightly different approach: we do not impose the fit to 
just one elemental ratio but we choose the mass cut that provides the best-fit to the
the elemental ratios measured on the surface of SDSS J102915+172927. 
From a technical point of view, since we inject energy in the star by imparting an arbitrary velocity 
$v_0$ to an internal mass layer, the final size of the mass of the remnant will be controlled by this 
(artificial) parameter (the larger $v_0$ the smaller the mass of the remnant).

Following this procedure, we find that the best overall fit to the observations is achieved for progenitor masses 
of $\rm 20~M_\odot$ and $\rm 35~M_\odot$ with mass cuts, explosion energies and ejected $\rm ^{56}Ni$ mass equal to:
$\rm 1.73~M_\odot$, $\rm 10^{51} erg$, $\rm M(^{56}Ni)=0.06~M_\odot$ and $\rm 1.73~M_\odot$, $\rm 2.4 \times 10^{51} erg$, 
$\rm M(^{56}Ni)=0.16~M_\odot$, respectively. These results represent standard values for ordinary core-collapse SN, in
agreement with what has been recently found by Heger \& Woosley (2010) to interpret the observed abundances of 
EMPs with typical Galactic halo signatures. Figure 1 shows the comparison between SDSS J102915+172927 elemental ratios 
relative to iron and the chemical yields of Pop III SN with 
$\rm 20~M_{\odot}, 35~M_{\odot}$, and $\rm 50~M_{\odot}$ progenitor masses. 
Except for the [Ti/Fe] and [Ni/Fe] ratios, 
the chemical yields of the $\rm 20~M_{\odot}$ and $\rm 35~M_{\odot}$ 
SN models provide a good fit to the observations. It is well known that the fit to the [Ti/Fe] and [Ni/Fe] abundance 
ratios can be improved by assuming some kind of unconventional behavior of the deep interior of the star, like 
the mixing and fallback mechanism proposed by Umeda \& Nomoto (2005). We stress, however, that these uncertainties 
do not affect the computation of dust production in the SN ejecta, which is the main purpose of the present investigation.

\begin{figure}
\center{\epsfig{figure=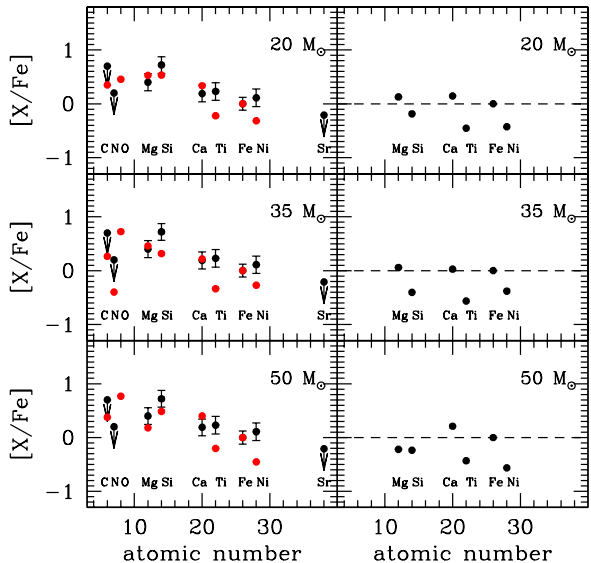, height=9.5cm}}
\caption{Comparison between the observed element abundance ratios of
the star SDSS J102915+172927 and the chemical yields of Pop III SN 
with progenitor masses of $\rm 25~M_{\odot}, 35~M_{\odot}$, 
and $\rm 50~M_{\odot}$ (from top to bottom). The left panels show the 
observationally inferred abundance ratios (black dots with errorbars or upper limits)
compared to the Pop III SN yields (red dots). The right panels show the corresponding 
residuals computed only for the elements with a real detection (excluding the upper limits).}
\end{figure}

\section{Dust yields from the metal free SN progenitors}

As the SN expands and cools, the metal-rich ejecta enters a temperature and density
regime where dust condensation can occur. We use the output of the SN explosion simulations
to compute the nucleation and accretion of dust species with a previously developed code that
has been applied to core-collapse (Todini \& Ferrara 2001; Bianchi \& Schneider 2007) and 
pair-instability SNe (Schneider et al. 2004) and that can
successfully reproduce observational data of SNe (Matsuura et al. 2011) and young SN remnants (Rho et al. 2008).
The calculation is based on classical nucleation theory: when a gas becomes supersaturated,
particles (monomers) aggregate in a seed cluster which subsequently grows by accretion of other monomers.
The dust mass and composition predicted for the two SN progenitors are shown in Figure 2.
We show the results obtained assuming that the seed clusters are formed by a minimum of 2 monomers and that
the sticking coefficient (the probability that an atom colliding with a grain will stick to it) is equal to 1 for
all grain species.

The differences among the two SN models reflect variations in the ejecta chemical composition, and - to a larger extent -
in the thermal evolution and expansion velocity. As a result of the hot and compact structure characterizing
the Pop III SN progenitors, dust condensation occurs when the nucleation current (the formation rate
of seed grains per unit volume) is large and relatively
small grains are formed (Bianchi \& Schneider 2007). The nucleation and accretion processes result in a typical Log-normal grain
size distribution with $\sim 0.01~\mu$m silicates and $\sim 5\times 10^{-3}~\mu$m magnetite grains
dominating the mass budget for the 20 and 35 $\rm M_{\odot}$ SN models, respectively.

The shocked ambient gas drives a reverse shock in the ejecta, which, by about 1000 years, has swept over a
considerable fraction of its volume. The passage of the reverse shock can be particularly destructive for
the small dust grains, due to the transfer of thermal and kinetic energy during impact with gas particles
(sputtering). If the SN explodes in a denser circumstellar medium, the reverse shock travels faster inside the
ejecta and encounters a gas at higher density. This increases the effect of sputtering. When the number of monomers
in each grain becomes less than 2, the grain is evaporated returning the metals into the gas phase.
Depending on the average density of the circumstellar medium, of the original
$\rm \sim 0.4~M_{\odot}$ {\rm of dust} formed in the ejecta of the two SN models, between 1 and 20\% is able to survive
the passage of the reverse shock. A large fraction of newly formed dust mass is returned to the gas phase
on a time scale of $\sim [3 - 20] \times 10^3$~yr after the explosion, thus fixing the final depletion factor $\rm f_{dep} = 
M_{d}/(M_{d}+M_{Z})$, where $\rm M_{d}$ ($\rm M_{Z}$) is the dust (metal) mass.
The dust mass is dominated by $\rm Mg_2SiO_4$ and $\rm Fe_3O_4$ grains for the
$\rm 20~M_{\odot}$ and the $\rm 35~M_{\odot}$  SN models, respectively.
These remain the dominant dust species after the partial destruction in the reverse shock (see Figure 2).

\begin{figure}
\center{\epsfig{figure=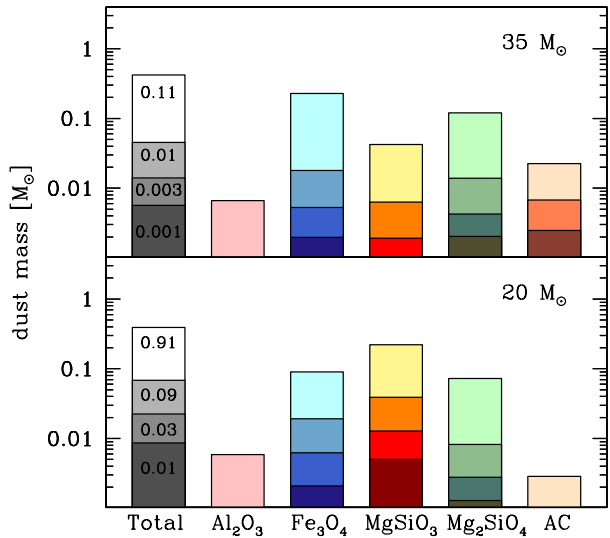, height=9.5cm}}
\caption{Dust mass and composition for the two SN progenitors. 
Each histogram represents the mass in $\rm M_{\odot}$ formed at the end of the condensation phase
(top light colour) and after the passage of the reverse shock assuming three different densities of the
circumstellar medium: $\rho = 10^{-25}$, $10^{-24}$ and $10^{-23}$~g cm$^{-3}$ from top to bottom; the first histogram
represents the total mass of dust; numbers indicate the value of the depletion factor $\rm f_{\rm dep}$.
Other histograms indicate the mass contribution of each dust species.}
\end{figure}

\section{Birth conditions of SDSS J102915+172927}

According to the recent analysis by Schneider et al. (2011),
in the extremely metal-poor environments, such as the birthplace of SDSS J102915+172927,
sub-solar mass stars can form only if,
\[
{\cal D} > 1.4 \times 10^{-8} \left[\frac{S}{10^5 \mbox{cm}^2\mbox{gr}^{-1}}\right]^{-1} 
\left[\frac{T}{10^3 \mbox{K}}\right]^{-1/2} \left[\frac{n_{\rm H}}{10^{12} \mbox{cm}^{-3}}\right]^{-1/2},
\]
\noindent
where ${\cal D} = Z \, \rm f_{dep}$ is the dust-to-gas ratio, $S$ is the grain geometric cross section per unit dust mass
and $n_{\rm H} = 10^{12}$~cm$^{-3}$, $T = 10^3$~K are the characteristic density and temperature at which dust
cooling becomes efficient.
At the observed metallicity of SDSS J102915+172927, this yields to $\rm f_{dep} > 0.01$,
when the values $S \sim [1.5 - 2.5] \times 10^5 \,\mbox{cm}^2\,\mbox{g}^{-1}$, appropriate for shock-processed
SN grains, have been adopted.
According to the results shown in Figure 2, if the parent gas cloud has been enriched by a metal-free
$\rm 20~M_{\odot}$ SN, the criterium is satisfied for three out of the four shock models explored,
i.e. if the SN explodes in a medium with density $\le 10^{-24}$~gr cm$^{-3}$.
If the enrichment has been caused by the explosion of a $\rm 35~M_{\odot}$ SN, due to the
smaller grain sizes, the fragmentation conditions are met only if no destruction by the
reverse shock occurs, i.e. if the explosion takes place in a very rarefied medium.

These results have been confirmed by means of a semi-analytical model
that follows the thermal and chemical evolution during the collapse,
including a detailed description of all the relevant molecular, metal line, and dust
grain cooling processes (Figure 3).
A full description of the semi-analytical model can be found in published
papers (Omukai et al. 2005; Schneider et al. 2006, 2010).
Gravitational fragmentation occurs when $(\partial \log T/\partial \log n_H) < 0$;
the minimum fragment mass corresponds to the Jeans mass at the inflection point of
the equation of state (Larson 2003).
Depending on the metallicity, the first phase of fragmentation is due to molecular or metal line cooling.
For all the models presented in Figure 3, we find that at $n_{\rm H} \sim 10^5 - 10^9$~cm$^{-3}$, cooling
is dominated by OH molecules. The minimum fragment that form have a mass $\rm \simeq 8~M_{\odot}$.
When the density increases to $n_{\rm H} \ge 10^{10}$~cm$^{-3}$, dust continuum emission starts to be efficient and the gas
enters a new phase of cooling; where the fragmentation conditions are met (Schneider \& Omukai 2010), 
the typical masses that form are $\rm \simeq [0.05 - 0.1]~M_{\odot}$.
Although these numbers can not be directly associated to the final stellar masses,
theoretical arguments, observational data and numerical simulations confirm that the thermal properties
of star-forming clouds play an important role in determing the stellar initial mass function,
and support the hypothesis that the Jeans mass at the point of minimum temperature
is important in determing the stellar characteristic mass (Larson 2003).

\begin{figure}
\center{\epsfig{figure=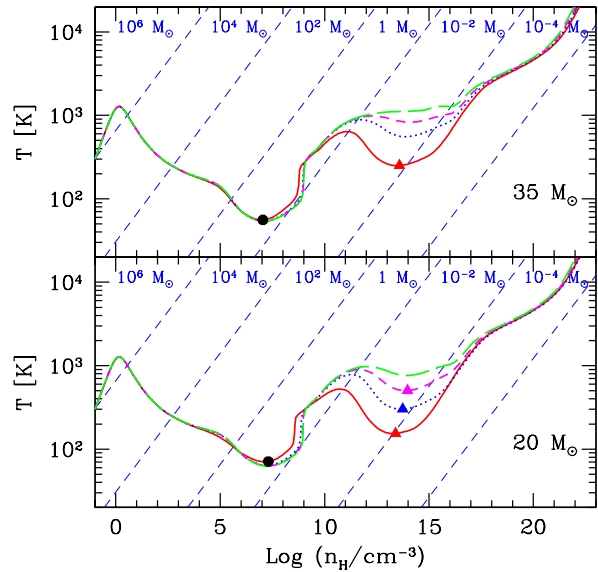, height=9.5cm}}
\caption{Thermal evolution during the collapse of star forming clouds enriched by the products of metal-free
$\rm 20~M_{\odot}$ (bottom panel) and $\rm 35~M_{\odot}$ (top-panel) SN explosions.
We fix the total metallicity (including metals and dust grains) to be $\rm Z = 4.5 \times 10^{-5} Z_{\odot}$, so as to reproduce
the birth conditions of SDSS J102915+172927.
Each track corresponds to a specific value of the dust depletion factor shown in Figure 2,
decreasing from the bottom to the top curves.
The points with dots and triangles mark the states where fragmentation conditions are met.
We identify the fragmentation epochs by requiring that the adiabatic index $\gamma$ becomes
greater than 1 after a phase of cooling (where $\gamma < 1$, see Schneider \& Omukai 2010).
The diagonal dotted lines represent constant Jeans mass values in the $(n_{\rm H}, T)$ plane.}
\end{figure}

\section{Discussion and conclusions}
The formation of the first low-mass stars requires the presence in the star-forming gas of 
metals, mostly C and O, which contribute to gas cooling. 
In the ejecta of core-collapse supernovae, a fraction of these metals 
can condense into dust grains that provide an additional cooling channel to the star forming gas. 
The relative importance of metals and dust grains in the formation of the first low-mass
stars is still a subject of debate. 

Observations show that the fraction of CEMPs in the Galactic halo increases with decreasing metallicity and 
that three out of the four halo stars with [Fe/H] $<$ -4 are carbon-enhanced. This has given support
to a metal (mostly O and C)-driven transition in the fragmentation scales that allows the formation
of low-mass stars when [C/H] and [O/H] (or a combination of the two) exceed a critical threshold
(Bromm \& Loeb 2003; Frebel et al. 2007). Despite the fact that the origin of the large CNO  
enhancement in CEMPs is still unclear (see Section 2), if these elements were already present in
the parent birth clouds, the formation of low-mass CEMPs does not appear to be problematic, even
at the lowest observed [Fe/H] (Schneider et al. 2003). 

The recent discovery of SDSS J102915+172927 challenges this scenario because the observed 
[C/H] and [O/H] places this star in a 'forbidden-zone' for low-mass star formation via 
metal-fine structure line cooling.
 Our analysis shows that the formation of sub-solar mass 
stars at very low metallicities, such as the one inferred for SDSS J102915, relies on dust.
The observed properties of this primeval star constrain the supernova progenitors
to have masses $\rm \sim [20 - 40]~M_{\odot}$ and to release $\rm \sim [0.01 - 0.4]~M_{\odot}$
of dust in the surrounding medium. If these values are representative of the properties of the first
supernovae, then rapid dust enrichment must have followed the formation of the first stars.
As a consequence, the transition from a star formation epoch dominated by massive Population III
stars to an epoch where Population II and I stars form with an ordinary stellar initial mass function
may be driven by dust and proceed more rapidly than previously thought (Tornatore et al. 2007; Maio et al. 2010).
This has important implications for the nature of the first galaxies and
their contribution to the process of cosmic reionization.


\label{lastpage}

\end{document}